\documentclass{article}
\usepackage{spconf,amsmath,graphicx,hyperref}
\usepackage{placeins} 
\usepackage{subfigure} 

\title{MoA-Off: Adaptive Heterogeneous Modality-Aware Offloading with Edge-Cloud Collaboration for Efficient Multimodal LLM Inference}
\name{Zheming Yang\textsuperscript{\rm 1}, Qi Guo\textsuperscript{\rm 1,3}, Yunqing Hu\textsuperscript{\rm 1,3}, Chang Zhao\textsuperscript{\rm 1,3}, Chang Zhang\textsuperscript{\rm 1,3}, Jian Zhao\textsuperscript{\rm 1,3,4}, Wen Ji\textsuperscript{\rm 1,2,4}}
\address{ \textsuperscript{\rm 1}Institute of Computing Technology, Chinese Academy of Sciences, Beijing, China \\ \textsuperscript{\rm 2}Institute of AI for Industries, Nanjing, China  \\ \textsuperscript{\rm 3}University of Chinese Academy of Sciences, Beijing, China   \\ \textsuperscript{\rm 4}Peng Cheng Laboratory, Shenzhen, China}

\usepackage{color}

\begin{document}
%
\maketitle
\begin{abstract}
Multimodal large language models (MLLMs) enable powerful cross-modal inference but impose significant computational and latency burdens, posing severe challenges for deployment in resource-constrained environments. In this paper, we propose MoA-Off, an adaptive heterogeneous modality-aware offloading framework with edge-cloud collaboration for efficient MLLM inference. MoA-Off introduces a lightweight heterogeneous modality-aware module that estimates the complexity of heterogeneous inputs through multi-dimensional feature analysis. Then, an adaptive edge-cloud collaborative offloading strategy is proposed that dynamically schedules workloads between edge and cloud based on modality-aware complexity scores and real-time system states. The experimental results demonstrate that MoA-Off can achieve over 30\% reduction in latency and 30\%–65\% decrease in resource overhead while maintaining competitive accuracy compared to traditional approaches.

\end{abstract}


\section{Introduction}
\label{sec:intro}
\begin{figure}[t]
  \centering
   \includegraphics[width=1\columnwidth]{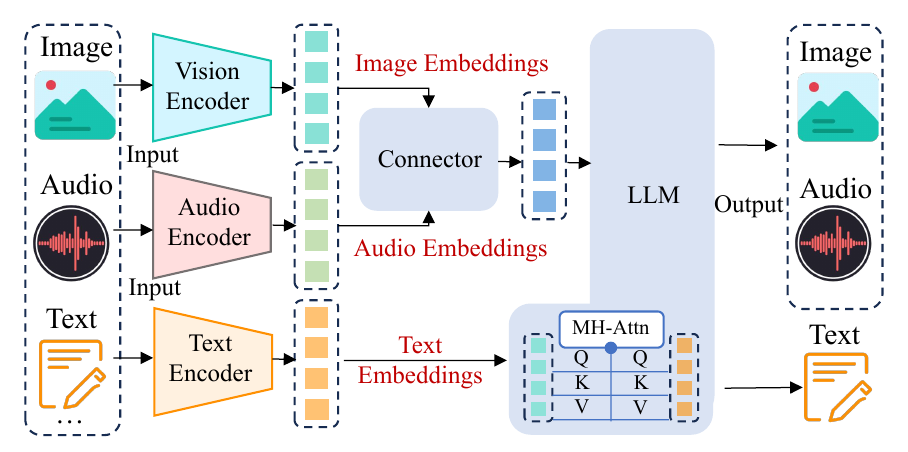}
  \caption{The examples of MLLM.}
  \label{figure1}
\end{figure}

Multimodal large language models (MLLMs) have recently demonstrated remarkable capabilities in integrating visual, auditory, and textual information, enabling unified reasoning across heterogeneous data sources \cite{wu2023multimodal111}. as shown in Fig.~\ref{figure1}. These models are increasingly deployed in applications ranging from intelligent assistants to real-time perception systems \cite{koh2023generating222}. However, their impressive performance comes at the cost of substantial computational and memory demands \cite{lin2025boosting333}, which pose severe challenges in latency-sensitive and resource-constrained environments \cite{chen2024tomGPTXXX}. Pure cloud-based solutions, while powerful, often suffer from excessive transmission latency, bandwidth overhead, and potential privacy risks \cite{zhang2025gsmm444}. In contrast, edge-only approaches provide lower response time and better locality but are fundamentally limited by restricted computing capacity\cite{yao2025efficient555}, leading to degraded performance on complex multimodal tasks \cite{zheng2025review666}.

To address these limitations, recent research has explored edge–cloud collaborative inference \cite{yang2025ec2moe777,wang2024cloud888}, in which lightweight pre-processing is conducted on edge devices and computation-intensive reasoning is offloaded to the cloud. Such hybrid solutions balance efficiency and accuracy by distributing workloads across heterogeneous resources. Nevertheless, existing collaborative frameworks typically adopt uniform offloading policies, treating multimodal inputs without considering their intrinsic heterogeneity \cite{yang2024perllm101010}. In practice, different modalities exhibit distinct computational characteristics, communication costs, and sensitivity to inference accuracy \cite{xie2024advancing111111}. Ignoring these differences often results in suboptimal allocation of tasks, inefficient resource utilization, and inconsistent user experience \cite{yi2025enhancing121212}. 


In this paper, we present MoA-Off, an adaptive modality-aware offloading framework for efficient MLLM inference with edge–cloud collaboration. Our approach introduces modality-specific complexity into the offloading process, thereby improving both system efficiency and inference performance. The main contributions are as follows:

\begin{itemize}
\item We develop a lightweight heterogeneous modality-aware module, which estimates the complexity of image and text inputs through efficient feature extraction, while maintaining minimal overhead on edge devices. 
    
\item We propose an adaptive edge-cloud collaborative offloading policy. This policy enables fine-grained scheduling, ensuring that resource-intensive modalities are offloaded to the cloud for accurate reasoning, while lightweight inputs are retained on the edge for low-latency processing.
    
\item We conduct extensive experiments on multimodal benchmark datasets. The experimental results demonstrate that MoA-Off achieves over 30\% reduction in latency while maintaining competitive accuracy, alongside a 30\%–65\% decrease in resource overhead,  compared to traditional approaches.
\end{itemize}

\section{Preliminaries}
\label{sec:format}
\textbf{Multimodal LLM.} MLLM extends conventional text-based LLMs by integrating heterogeneous modalities such as text, images, audio, and video into a unified reasoning framework \cite{yin2024survey202020}. Formally, given an input set of modalities ${x_t, x_v, x_a}$ corresponding to text, vision, and audio, the model constructs modality embeddings $h_m = f_m(x_m)$ through respective encoders $f_m(\cdot)$, and then aligns them into a common latent space via a projection function $g(\cdot)$:
\begin{equation}
z_m=g\left(h_m\right), \quad m \in\{t, v, a\} .
\end{equation}

This formulation empowers MLLMs to handle complex tasks such as multimodal question answering, captioning, and grounded reasoning, yet also results in significant computational and memory demands, making efficient deployment a non-trivial challenge.

\textbf{Edge-Cloud Architecture.} Edge-cloud architectures provide a hierarchical computing paradigm that distributes inference workloads across resource-constrained edge devices and powerful cloud servers, thereby balancing latency, energy efficiency, and computational capacity \cite{yang2021intelligent212121}. This division enables low-latency responses while leveraging the scalability of the cloud for complex reasoning \cite{tong2016hierarchical131313}. Such a collaborative design mitigates the limitations of standalone edge inference while reducing cloud dependency compared to purely centralized solutions, making it especially suitable for real-time multimodal applications under dynamic network and resource conditions.

\section{The Proposed Method}
\label{sec:method}
\begin{figure}[t]
  \centering
 \includegraphics[width=1\columnwidth]{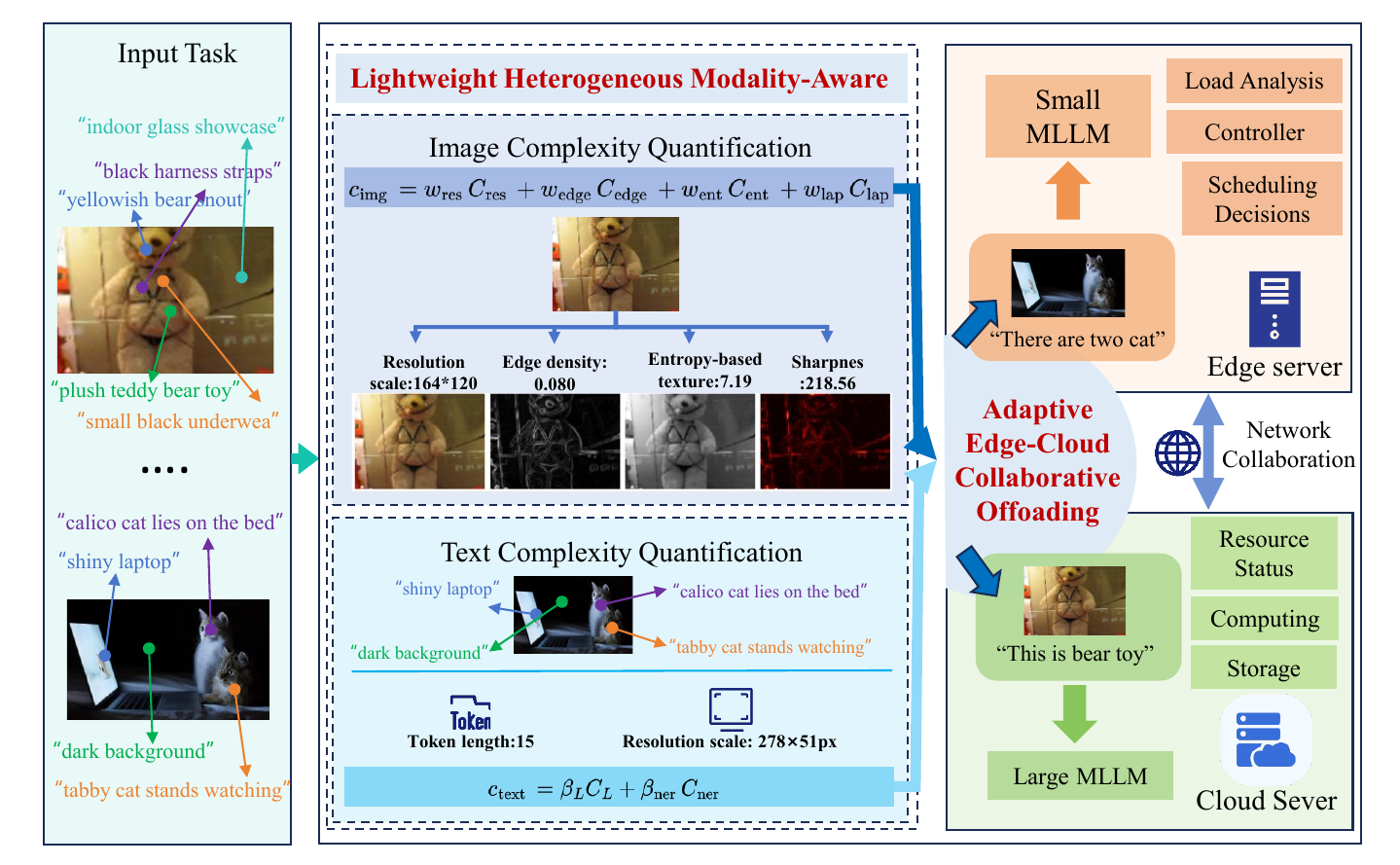}
  \caption{The overview of the proposed MoA-Off framework.}
  \label{figure2}
  \vspace{-2mm}
\end{figure}

We propose MoA-Off, an adaptive heterogeneous modality-aware offloading framework designed to enhance the efficiency of MLLM inference by edge–cloud collaboration, as illustrated in Fig.~\ref{figure2}. It primarily consists of two parts: (1) \textit{Lightweight Heterogeneous Modality-Aware}, and (2) \textit{Adaptive Edge-Cloud Collaborative Offloading}. The core idea of MoA-Off is to exploit the heterogeneous characteristics of different modalities and dynamically allocate computational workloads between edge and cloud. 


\subsection{Lightweight Heterogeneous Modality-Aware}
The lightweight heterogeneous modality-aware module is designed to efficiently identify the type and estimate the complexity of each input modality on the edge device. Its primary objective is to provide lightweight yet reliable complexity scores, which serve as the basis for the adaptive offloading policy. To ensure practicality, all metrics are derived from fast, single-pass feature extraction techniques \cite{yang2023javp151515}, making them suitable for real-time deployment in resource-constrained environments.

\subsubsection{Image Complexity}
For an input image $I \in [0,255]^{H \times W}$ with $N=HW$ pixels, the image complexity is quantified as a weighted combination of several lightweight indicators that capture different aspects of visual information. Specifically, we consider resolution scale, edge density, entropy-based texture, and sharpness. The overall complexity is computed as a weighted sum: $c_{\mathrm{img}}=w_{\mathrm{res}} C_{\mathrm{res}}+w_{\mathrm{edge}} C_{\mathrm{edge}}+w_{\mathrm{ent}} C_{\mathrm{ent}}+w_{\mathrm{lap}} C_{\mathrm{lap}}$, where the weights $w$ are non-negative and normalized to sum to one. Each indicator is defined as follows.

\textbf{Resolution scale.}
Higher-resolution images typically require more computation. We normalize image size relative to a reference resolution $(H_0,W_0)$: $C_{\mathrm{res}}=\min \left(1, \frac{H W}{H_0 W_0}\right)$.

\textbf{Edge density.}
Images with dense edges usually contain richer structural information and thus higher processing costs. We measure the average Sobel gradient magnitude and normalize it:
\begin{equation}
C_{\text {edge }}=\operatorname{clip}\left(\frac{\bar{G}-P_5(G)}{P_{95}(G)-P_5(G)+\epsilon}, 0,1\right) .
\end{equation}


\textbf{Entropy-based texture.}
Texture richness is approximated by image entropy, derived from the gray-level histogram:
\begin{equation}
H(I)=-\sum_{k=0}^{255} p_k \log p_k, \quad C_{\mathrm{ent}}=\frac{H(I)}{\log 256},
\end{equation}
where $p_k$ is the normalized frequency of gray-level $k$.

\textbf{Sharpness.}
Sharp or detailed images are more demanding to process. We use the variance of the Laplacian operator as a sharpness indicator:
\begin{equation}
C_{\text {lap }}=\operatorname{clip}\left(\frac{\operatorname{Var}\left(\nabla^2 I\right)-P_5}{P_{95}-P_5+\epsilon}, 0,1\right),
\end{equation}
where $P_{5}$ and $P_{95}$ represent the $5$-th and $95$-th percentiles of the Laplacian variance across a calibration set.





\subsubsection{Text Complexity}
Text complexity reflects the processing overhead introduced by long sequences, entity density, and context dependency \cite{bae2023complexitynet191919}.
The final text complexity score is computed as: $c_{\text {text }}=\beta_L C_L+\beta_{\text {ner }} C_{\text {ner }}$, where $\beta_L$ and $\beta_{\text{ner}}$ are weights.

\textbf{Token length.}
Longer sequences lead to quadratic growth in transformer computation. We normalize token count $L$ relative to a threshold $L_0$: $C_L=\min \left(1, \frac{L}{L_0}\right)$.

\textbf{Entity or numeric density.}
The presence of named entities, numerical expressions, or specialized terminology often increases semantic reasoning complexity. We measure this factor by computing the average number of entities per sentence: $C_{\mathrm{ner}}=\min \left(1, \frac{E / S}{\gamma}\right)$, where $E$ is the number of entities and $S$ is the number of sentences, $\gamma$ is a scaling constant.

\subsection{Adaptive Edge-Cloud Collaborative Offloading}
\label{subsec:selector}
Building upon the modality-specific complexity estimation introduced in \textit{Section 3.1}, we design an adaptive offloading strategy that dynamically determines whether each modality should be processed locally at the edge or offloaded to the cloud. The key objective is to minimize latency and resource overhead while maintaining inference accuracy. 


For each modality $m_i \in {\text{text}, \text{image}}$, let $c_i \in [0,1]$ denote the estimated complexity score from the modality perception module. The system state is represented by $s=(\ell,b)$, where $\ell$ is the current edge load and $b$ is the available network bandwidth. The offloading decision $d_i$ is defined as:
\begin{equation}
d_i= \begin{cases}\text { edge }, & c_i \leq \tau_{m_i} \wedge \ell \leq \ell_{\max } \wedge b \leq \beta, \\ \text { cloud, } & \text { otherwise, }\end{cases}
\end{equation}
where $\tau_{m_i}$ is the modality-specific complexity threshold, $\ell_{\max}$ is the maximum tolerable edge utilization, and $\beta$ is a bandwidth limit.

In multimodal tasks, different modalities often exhibit divergent levels of complexity and resource demand. Instead of enforcing a uniform offloading decision for the entire input, our strategy employs a per-modality routing mechanism that enables partial offloading and fine-grained scheduling. For instance, in an image and text query, the image, characterized by high computational and bandwidth requirements, can be offloaded to the cloud, while the accompanying short text is processed locally on the edge for immediate response. Formally, the decision vector is defined as
\begin{equation}
d = \pi(c_1, c_2, \ldots, c_k, s) \in {\text{edge}, \text{cloud}}^k,
\end{equation}
where $k$ denotes the number of modalities in the input, $c_i$ is the complexity score of modality $i$, $s$ represents the system state (including edge load and network bandwidth), and $\pi(\cdot)$ is the adaptive scheduling policy function that integrates modality-aware thresholds with system-level dynamics.


\section{Experiments}
\label{sec:experiments}

\subsection{Experimental Setup}
\label{subsec:setup}
We utilize an NVIDIA A100 (40 GB) GPU as the cloud server and an NVIDIA 3090 (24 GB) as the edge device. Qwen2-VL-2B model is deployed on the edge, while Qwen-2.5-VL-7B model is deployed in the cloud \cite{bai2023qwen181818}. The evaluation datasets are VQAv2 \cite{goyal2017making161616} and MMBench \cite{liu2024mmbench171717}, we randomly selected 5,000 images from the validation set for testing. The network bandwidth is set to 200 Mbps, 300 Mbps, and 400 Mbps, respectively. The weights for image complexity and text complexity are set to their average values, with the complexity threshold set to 0.5. The baseline methods include (1) \textit{Cloud-only} (Qwen-2.5-VL-7B), (2) \textit{Edge-only} (Qwen2-VL-2B), and (3) \textit{PerLLM} (an edge-cloud collaboration solution). Evaluation metrics include accuracy, End-to-end latency, and resource overhead.

\subsection{Main Results and Analysis}
\label{subsec:main_result}

\subsubsection{Accuracy Comparison}
As shown in Table~\ref{tab1}, the proposed MoA-Off framework achieves highly competitive performance. Critically, it attains an overall accuracy that is comparable to the cloud-only approach with a negligible performance penalty of merely $\textless$0.4\%. Furthermore, our method substantially outperforms the edge-only and PerLLM by a considerable margin of over 4.8\%-16.8\% in absolute accuracy. These results highlight the effectiveness of the proposed  MoA-Off framework.

\begin{table}[h]
\centering
\caption{Accuracy (\%) comparison results.}
\label{tab1}
\resizebox{\linewidth}{!}{
\begin{tabular}{|l|c|c|c|c|}
\hline
\multicolumn{1}{|c|}{\textbf{Method}} & \textbf{Cloud-only} & \textbf{Edge-only} & \textbf{PerLLM} & \textbf{MoA-Off} \\ \hline
\multicolumn{5}{|c|}{\textit{VQAv2 Dataset}} \\ \hline
200 (Mbps) & 76.3 & 61.4 & 71.3 & 76.1\\ \hline
300 (Mbps) & 77.4 & 63.2 & 71.8 & 77.2\\ \hline
400 (Mbps)   & 77.8 & 63.5 & 72.4  & 77.5\\ \hline
\multicolumn{5}{|c|}{\textit{MMBench Dataset}} \\ \hline
200 (Mbps) & 75.6 & 58.4 & 68.3 & 75.2\\ \hline
300 (Mbps) & 76.1 & 60.1 & 69.2 & 75.9\\ \hline
400 (Mbps)   & 76.5 & 61.2 & 69.9  & 76.3\\ \hline
\end{tabular}
}
\end{table}

\subsubsection{End-to-End Latency Comparison}
\begin{figure}[th]
	\centering 
    \subfigure[VQAv2]{
    \label{Fig.sub.3.1}
    \includegraphics[width=0.22\textwidth]{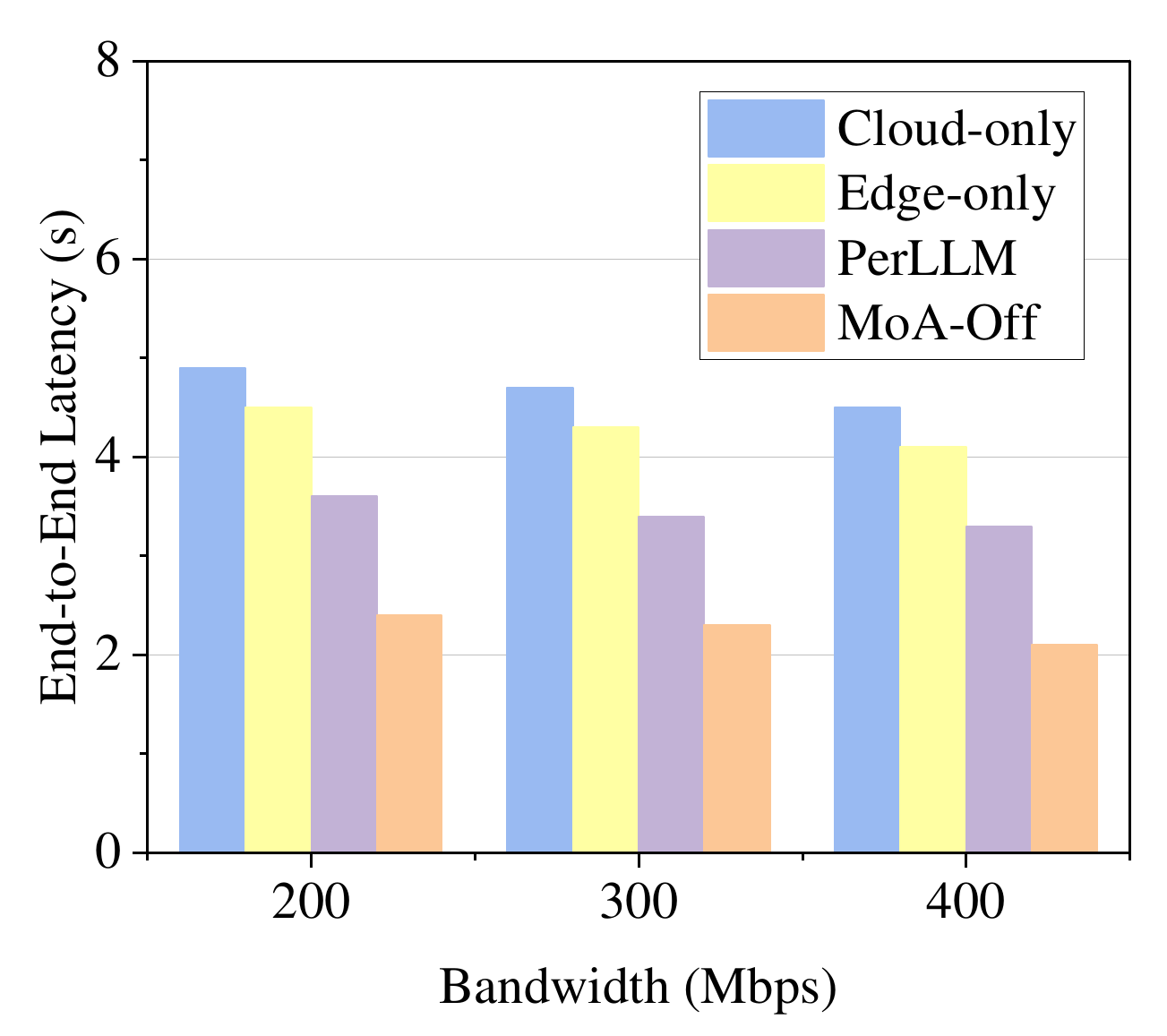}}
    \subfigure[MMBench ]{
    \label{Fig.sub.3.2}
    \includegraphics[width=0.22\textwidth]{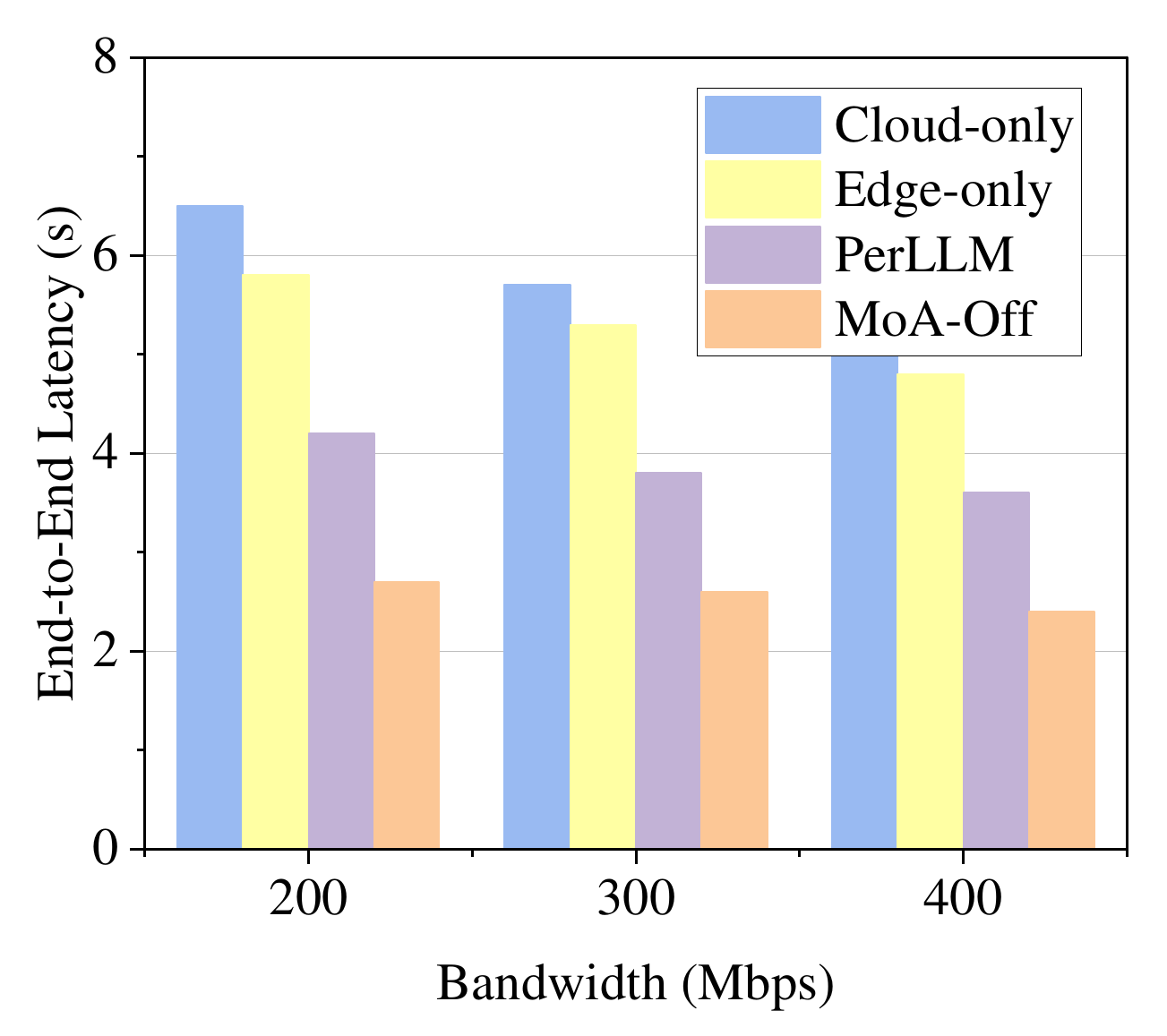}}
    
    \caption{The end-to-end latency comparison results of different methods. (a) shows results on the VQAv2 dataset, while (b) shows results on the MMBench dataset.}
	\label{figure3}
\end{figure}

 As shown in Fig.~\ref{figure3}, the comprehensive latency measurements reveal that our proposed MoA-Off framework achieves the lowest mean end-to-end latency among all evaluated strategies. It significantly outperforms the cloud-only approach and the edge-only approach by a margin of over 50\%. Compared to other edge-cloud collaborative methods, MoA-Off achieves over 30\% reduction in latency. Overall, by intelligently offloading only the most complex inferences to the cloud, our method avoids the severe latency tail typical of edge-only models struggling with difficult samples, thereby achieving an optimal balance.

\subsubsection{Resource Overhead Comparison}

\begin{figure}[t!]
	\centering 
    \subfigure[VQAv2, Computing]{
    \label{Fig.sub.4.1}
    \includegraphics[width=0.23\textwidth]{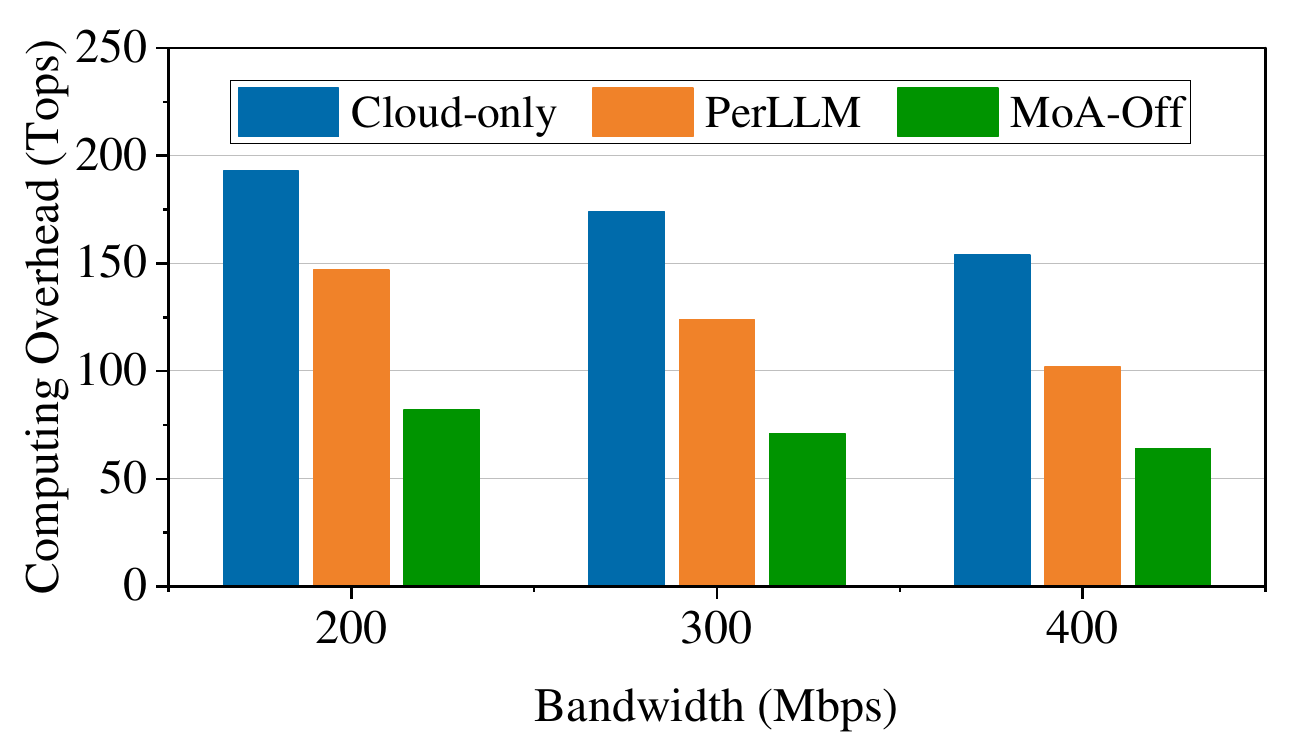}}
    \subfigure[MMBench, Computing   ]{
    \label{Fig.sub.4.2}
    \includegraphics[width=0.23\textwidth]{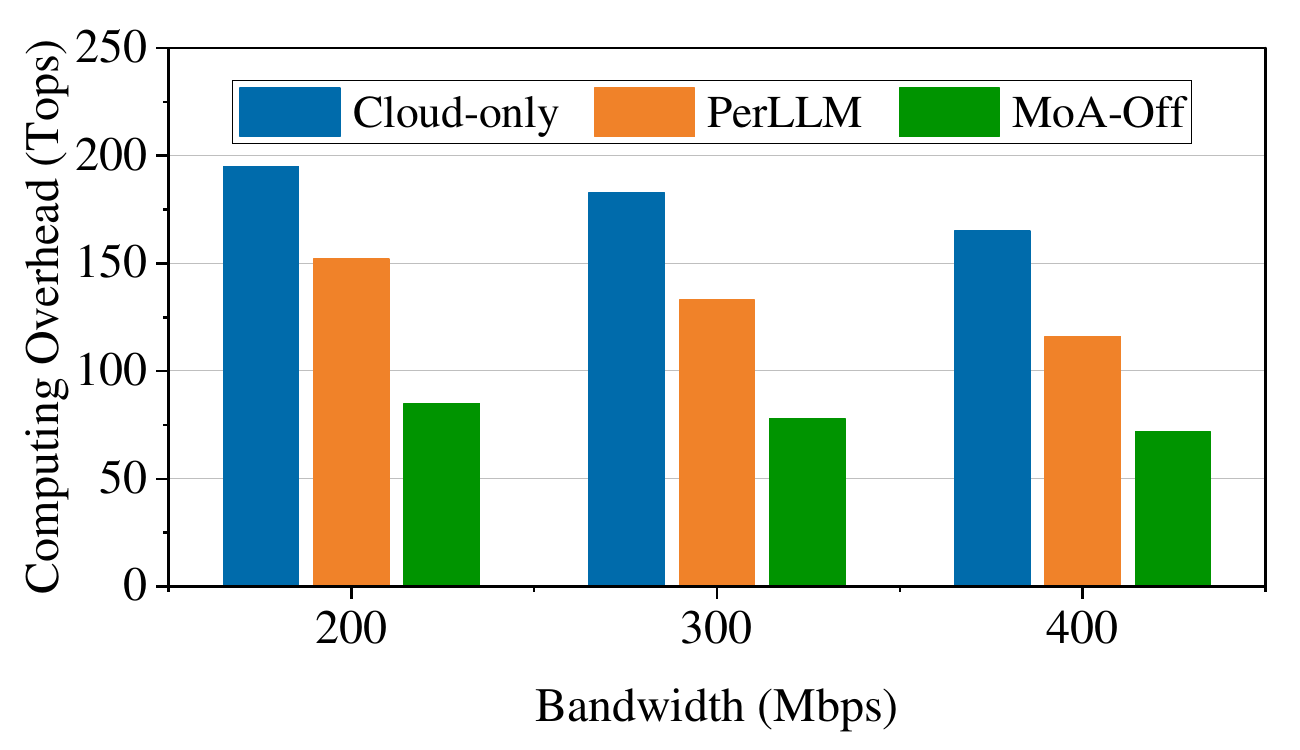}}

        \subfigure[VQAv2, Memory]{
    \label{Fig.sub.4.3}
    \includegraphics[width=0.23\textwidth]{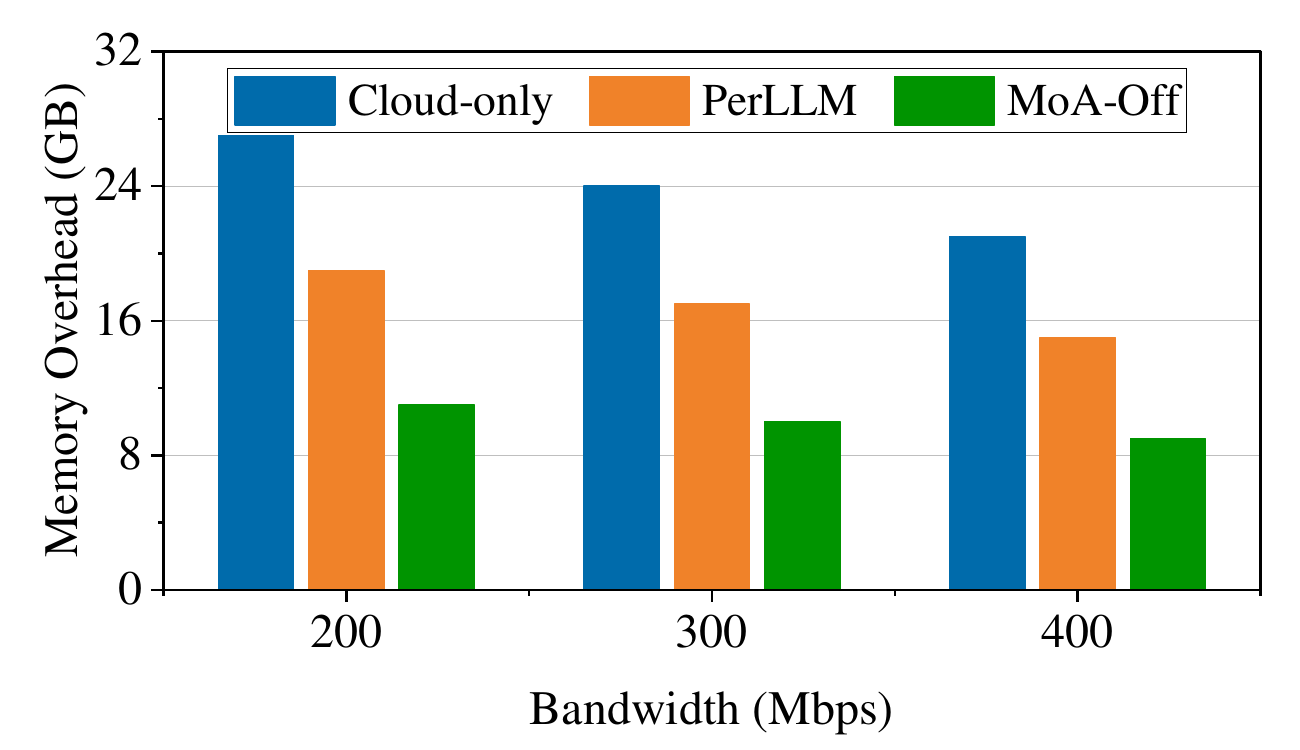}}
    \subfigure[MMBench, Memory]{
    \label{Fig.sub.4.4}
    \includegraphics[width=0.23\textwidth]{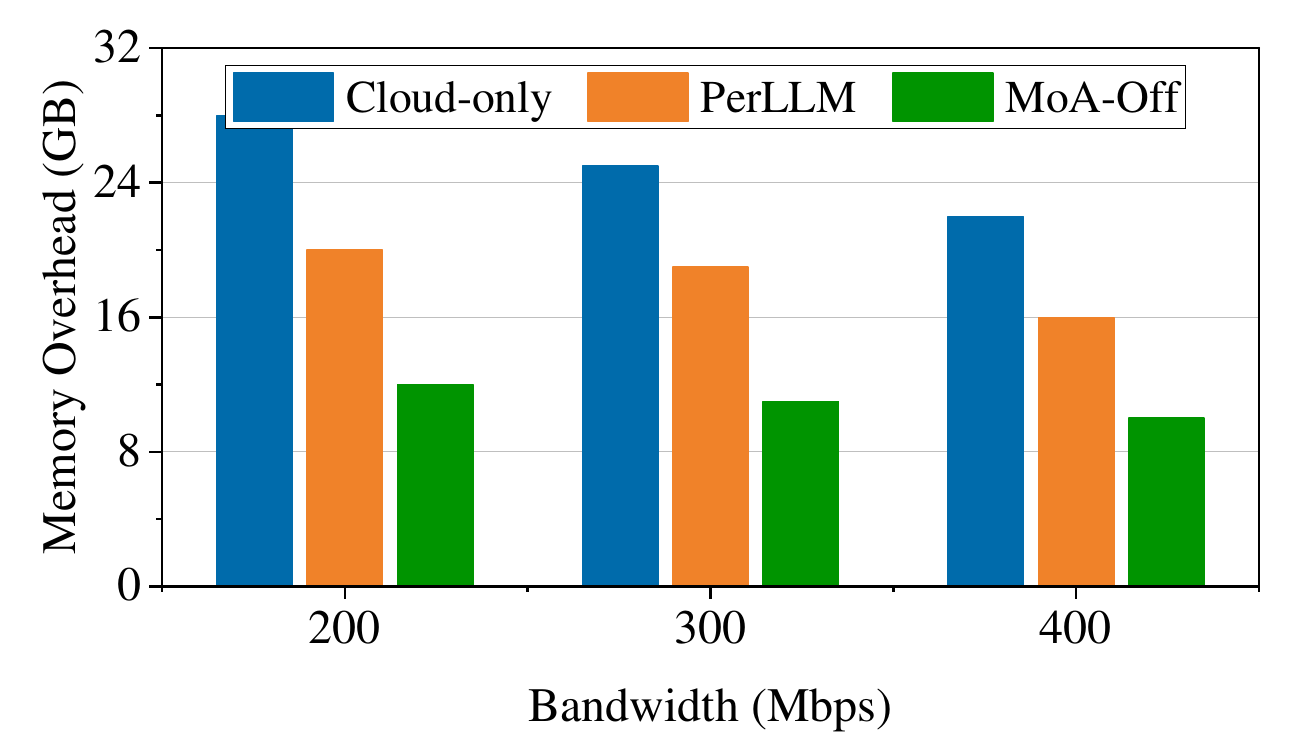}}
    
    \caption{The resource overhead comparison results of different methods. (a) and (b) represent computing overhead, while (c) and (d) represent memory overhead.}
	\label{figure4}
\end{figure}

\textbf{Computing Overhead.} As shown in Fig.~\ref{Fig.sub.4.1} and Fig.~\ref{Fig.sub.4.2}, MoA-Off exhibits the lowest computing overhead on both the VQAv2 dataset and the MMBench dataset. It drastically reduces the computational burden on the cloud infrastructure by offloading only the most complex samples. Compared to the cloud-only solution and PerLLM, MoA-Off can reduce computing overhead by 30\%-65\%. Our method introduces a negligible computing overhead on the edge device for calculating the modality-aware offloading policy, which is orders of magnitude lighter than running the MLMM. This strategic distribution of workload ensures that the vast majority of requests are resolved with minimal compute investment, either on the edge for simple tasks or on the cloud for complex ones.

\textbf{Memory Overhead.} As shown in Fig.~\ref{Fig.sub.4.3} and Fig.~\ref{Fig.sub.4.4}, by introducing adaptive modality-aware offloading, MoA-Off minimizes unnecessary data duplication and memory retention, effectively balancing allocation between edge and cloud. The cloud-only solution, benefiting from abundant resources, incurs substantial memory costs due to centralized multimodal model execution and frequent context reloading. In contrast, PerLLM alleviates part of this burden by distributing inference across the hierarchy, but its lack of modality-aware scheduling leads to redundant memory consumption at both ends, especially when handling heterogeneous modalities with varying complexity. 

\subsection{Ablation Study}
To further validate the effectiveness of our design, we performed ablation studies by removing the adaptive modality-aware offloading and edge–cloud collaborative scheduling components. Without the modality-aware offloading mechanism, inference accuracy dropped by 6.8\% on average, reflecting the system’s inability to account for heterogeneous modality complexity. Similarly, disabling the collaborative scheduling strategy increased end-to-end latency by 21.5\%, while computational and memory overhead rose by 18.7\% and 16.3\%, respectively, due to inefficient task distribution across edge and cloud. These results clearly demonstrate that both adaptive modality-aware offloading and collaborative scheduling are indispensable for efficient MLLM inference under resource-constrained edge–cloud environments.


\section{Conclusion}
In this work, we present MoA-Off, an adaptive heterogeneous modality-aware offloading framework for efficient MLLM inference. By introducing a lightweight modality perception module and an adaptive offloading strategy, MoA-Off effectively captures the heterogeneous characteristics of different modalities and dynamically allocates workloads between edge and cloud resources. Experimental results demonstrated that the proposed framework achieves significant reductions in latency and resource overhead while maintaining competitive inference accuracy, outperforming conventional edge-only, cloud-only, and edge-cloud offloading methods.


\bibliographystyle{IEEEbib}
\bibliography{strings,refs}

\end{document}